\documentclass[reprint,amsmath,amssymb,aps,prd,showpacs]{revtex4-1} 
\usepackage[T2A]{fontenc}
\usepackage[english]{babel}

\usepackage{graphicx,color}
\definecolor{darkblue}{rgb}{0,0,0.7}
\definecolor{darkred}{rgb}{0.7,0,0}
\usepackage{epstopdf}

\newcommand{\de}[0]{\partial}
\renewcommand{\div}[0]{{\rm div}\;}
\DeclareMathOperator{\grad}{grad}
\DeclareMathOperator{\rot}{rot}
\DeclareMathOperator{\erf}{erf}

\begin{document}

\title{Spontaneous crystallization noise in mirrors of gravitational wave detectors}
\author{N. M. Kondratiev$^{1,2}$}
\author{V. B. Braginsky$^1$}
\author{S. P. Vyatchanin$^1$}
\author{M. L. Gorodetsky$^{1,2}$}
\email{gorm@hbar.phys.msu.ru}
\affiliation{$^1$Faculty of Physics, Moscow State University, Moscow 119991, Russia \\
$^2$ Russian Quantum Center, Skolkovo 143025, Russia}

\date{\today}

\begin{abstract}

Core optics components for high precision measurements are made of stable materials, having small optical and mechanical dissipation. The natural choice in many cases is glass, in particular fused silica. Glass is a solid amorphous state of material that couldn't become a crystal due to high viscosity. However thermodynamically or externally activated stimulated local processes of spontaneous crystallization (known as devitrification) are still possible. Being random, these processes can produce an additional noise, and influence the performance of such experiments as laser gravitational wave detection.

\end{abstract}

\maketitle

\numberwithin{equation}{section}

\section{Introduction}

High-precision measurements always face a lot of noises and instabilities. The LIGO project \cite{LIGO,A-LIGO} have to account many fundamental sources of fluctuations. 
Fluctuations of temperature, which are translated into displacement of the mirror's surface through thermal expansion (thermoelastic noise) \cite{BGVte,BGVteC} and change of optical path due to fluctuations of refraction index (thermorefractive noise) \cite{BGVtr} combine producing generalized thermo-optical noise \cite{Zoo,Evans}. Better known Brownian fluctuations causing displacement of the mirror's surface \cite{Harry,HarryB} and photoelastic effect \cite{MyThermalNoise} produced by these fluctuations form the Brownian branch of noises. 

Not all of noise sources are easy to identify. In this work we are trying to estimate a noise coming from structural transformations in material. Fused silica is a glass -- neither crystal nor liquid. It is one of polymorphic forms of silicon dioxide and its internal energy is higher than that of crystalline state. The process of glass to solid crystallization at temperatures bellow glass transition is often called devitrification and was observed during long-term heating, under high-intensity laser exposition\cite{salleo2003laser} or ballistic impact\cite{Devitrification}. It is essential that different states have different material parameters (see table \ref{GCTab}), specifically density and refractive index.

\begin{table}[b]
\begin{center}
\begin{tabular}{|c|c|c|c|}
\hline
& Fused silica   & $\alpha$-quartz & Stishovite \\\hline
Density, g/cm$^3$							& $2.20$   & $2.65$					& $4.29$  \\\hline
Heat Capacity, J/(kg$\times$K)	& $1.052$  & $0.740$  			& $0.834$	\\\hline
Refractive index								& $1.46$ 	 & $1.54$ 				& $1.80$	\\\hline
\end{tabular}
\caption{\label{GCTab} Material parameters of polymorphic forms of silicon dioxide}
\end{center}
\end{table}

As one can see from the Table \ref{GCTab} the density of fused silica is smaller than that of crystalline quartz. In this way some contraction of fused silica samples in time should be observed if devitrification takes place. This effect of contraction is indeed known for glasses and its rate for different materials was measured \cite{DimStabFRiele,DimStabAkira,StabilitySilica}. From the other hand, each event of local crystallization is a discrete event causing small perturbations. The aim of the paper is to calculate the influence of possible local crystallization/reordering processes in the bulk of the matter on its surface and to calculate the spectral density of the surface fluctuations produced by this effect. We also give the empirical estimation of the rate at which such processes can happen in fused silica suspension fibers and use it to find the absolute values of the corresponding noise using the approach introduced in \cite{CreepNoise}.

\section{Noise of collapsing bubbles}

We start considering a piece of glass constituting the mirror, a small part  of which (which for simplicity we shall call a bubble) has changed it's state. This transition results in a local change of material parameters and also in equilibrium state parameters of such a bubble. One of these parameters is obviously the equilibrium volume, that is the volume of non-strained matter or exactly $m_{\rm b}/\bar{\rho}_{\rm c}$ (where $m_{\rm b}$ is the mass of the bubble, $\bar{\rho}_{\rm c}$ is the density of the crystal phase). But as a part of the bulk matter in the glassy state, this bubble still preserves the volume of the previous state's equilibrium volume $m_{\rm b}/\bar{\rho}_{\rm g}$, where is the density of glass.
In this way, from the difference of the equilibrium densities of the two phases we are getting an initial strain. 

We now have a binary system with the first component being  a crystal bubble with a deformation $u_b$
\begin{align}
\label{bubleEq}
&\ddot{\vec{u_b}}+c_{t_c}^2\rot\rot \vec u_b-c_{l_c}^2\grad\div\vec u_b=0,\\
&\vec u_b|_{t=0}=\vec u_0.
\end{align}
$c_{t_c}$ and $c_{l_c}$, consequently $c_{t_g}$ and $c_{l_g}$ are transversal and longitudinal speeds of sound in a crystalline or glassy state, $\vec u_0$ is time-independent initial deformation field.

\begin{figure}[h]
\center
\includegraphics[width=0.47\textwidth]{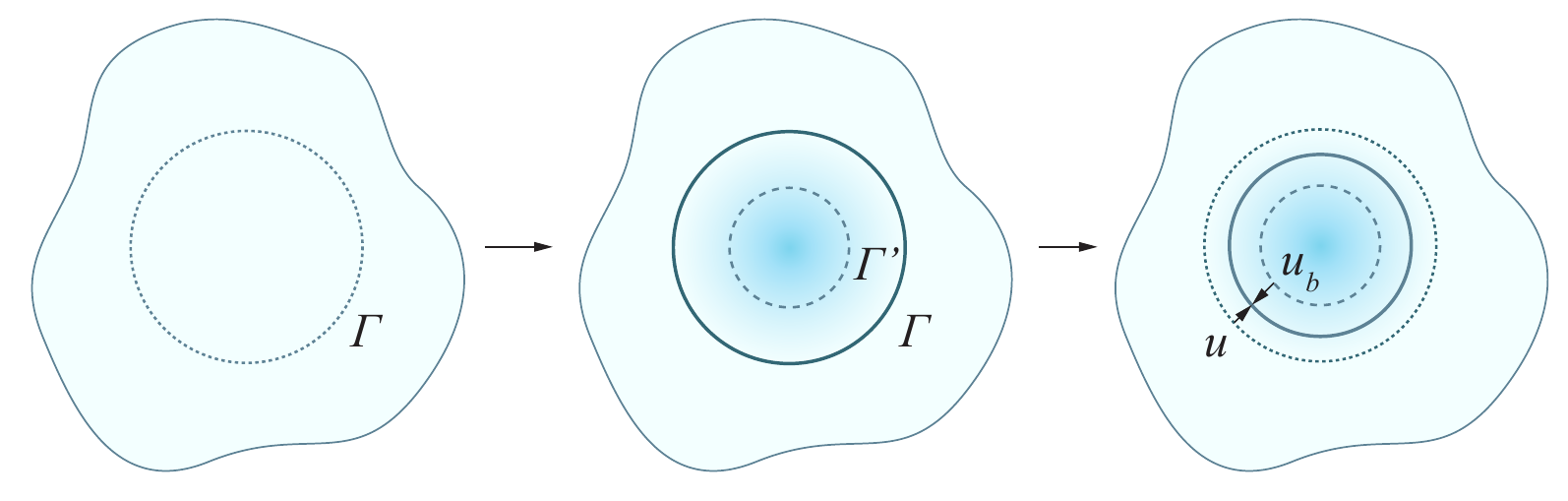}
\caption{Evolution of a phase transition bubble. The initial state bubble with radius $a$ (left). The bubble changes state to a phase with equilibrium radius $a'$ and gets strained (center). The system transforms to a new, deformed equilibrium state (right).}
\label{InitStrProblemPic}
\end{figure}

The second component is a glass. As a model task to understand the influence of such bubbles on the surface we are considering half-space in cylindrical coordinates~($\rho$,~$z$,~$\phi$).
\begin{align}
\ddot{\vec{u}}+c_{t_g}^2\rot\rot \vec u-c_{l_g}^2\grad\div\vec u=0 
\end{align}
\begin{align}
\vec u|_\Gamma&=&\vec u_b|_{\Gamma'}-\vec u_0|_{\Gamma'},\nonumber\\
\hat \sigma\vec n_\perp|_{\Gamma}&=&\hat \sigma_b\vec n'_\perp|_{\Gamma'},\nonumber\\
\sigma_{z\rho}|_{z=0}&=&0= \bar{\rho}_g c_{t_g}^2\left(\frac{\de u_\rho}{\de z}+\frac{\de u_z}{\de \rho}  \right),\nonumber\\
\sigma_{z\varphi}|_{z=0}&=&0=\bar{\rho}_g c_{t_g}^2\left(\frac{\de u_\varphi}{\de z}+\frac{1}{\rho}\frac{\de u_z}{\de \varphi}  \right),\nonumber\\
\sigma_{zz}|_{z=0}&=&0=2 \bar{\rho}_g c_{t_g}^2\frac{\de u_{z}}{\de z}+\nonumber\\
&+&(\bar{\rho}_g c_{l_g}^2-2\bar{\rho}_g c_{t_g}^2)\left(\frac{\de u_\rho}{\de \rho}+\frac{\de u_z}{\de z}+\frac{u_\rho}{\rho}  \right),
\end{align}
where $\hat\sigma$ is stress tensor, $\vec u$ is mirror deformation, $\Gamma$ is the initial boundary (equilibrium form of collapsing bubble of the first phase), $\Gamma'$ is equilibrium boundary of the second phase, $\vec n_\perp$ and $\vec n'_\perp$ are unit perpendiculars to those boundaries (see fig. \ref{InitStrProblemPic}).
In our case we assume the equilibrium states to be a sphere with the radius $a$ ($\Gamma\in[r=a]$) for the glass phase and $a'=a\sqrt[3]{\bar\rho_g/\bar\rho_c}$ ($\Gamma'\in[r=a']$) for the crystalline phase. The initial deformation can be obtained from the stationary spherical problem with displacement on the boundary:
\begin{align}
\label{lastEq}
&c_{t_c}^2\rot\rot \vec u_0-c_{l_c}^2\grad\div\vec u_0=0.\nonumber\\
&u_r|_{r=a'}=a-a'.
\end{align}

\section{Quasi-empirical estimation}
The above described stress-strain problem is still quite complex to solve analytically in arbitrary case. However, the initial displacement problem can be modeled by a thermoelastic problem, introducing thermal fluctuation in a bubble in the form
\begin{align}
T(\vec r)&=T_1 e^{-\frac{|\vec r -\vec r_0|^2}{b^2}}
\end{align}

\begin{figure}[h]
\center
\includegraphics[width=0.25\textwidth]{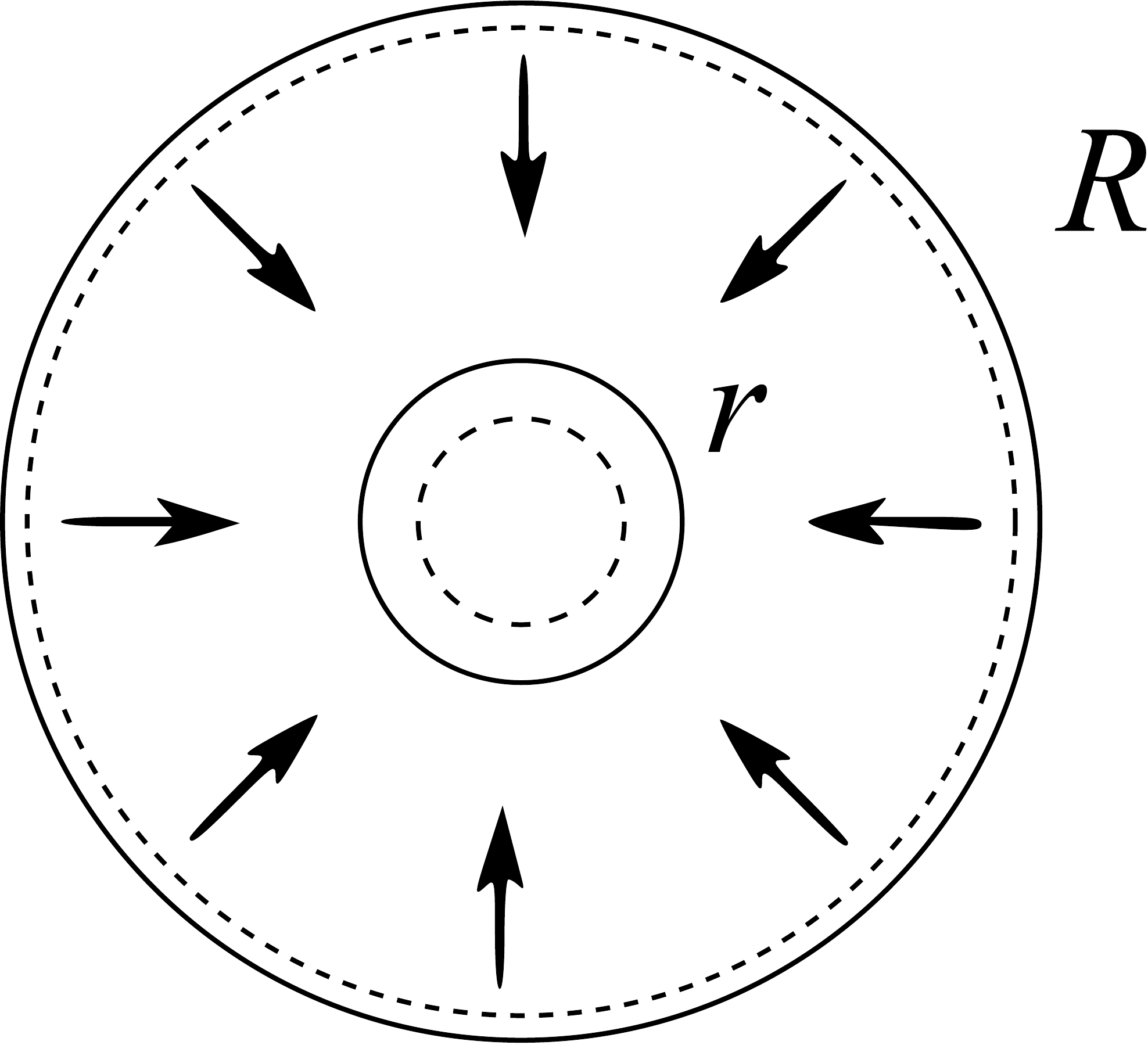}
\caption{Sphere radius change}
\label{Cylinder}
\end{figure}

Then the solution of the elastic equation with the heat source
\begin{align}
\label{ElastTeq}
\frac{1-\nu}{1+\nu}\grad\div\vec u-\frac{1-2\nu}{2(1+\nu)}\rot\rot\vec u =\alpha \grad T
\end{align}
will provide us a good estimate for \eqref{bubleEq}-\eqref{lastEq}, with local pressure substituted by a heat source ($\alpha$ is the coefficient of thermal expansion, $\nu$ is Poisson coefficient). To find the parameters $T_1$ and $b$ consider a pure spherical case (expansion of a spericall shell). In \cite{ll5en} we can find expressions for variations of inner and outer spherical radius change as shown on Fig. \ref{Cylinder}.
\begin{align}%
	\label{sphereForm}
	\frac{\delta r}{\delta R}= \frac{1}{3} \frac{1+\nu}{1-\nu} \frac{R^2}{r^2}.
\end{align}%
Here $R\gg r$ is assumed.
In our case $R$ is the distance of the collapsing bubble from the surface of the mirror, $\delta R$ is the surface displacement, $r$ is simply $a$ -- the radius of the collapsing bubble and $\delta r$ is $u_{\rm eq}$ -- the stationary solution of the system \eqref{bubleEq}-\eqref{lastEq}. For a rough estimation it can be approximated as
\begin{align}%
\label{deltar}
	\delta r\approx a\left(1-\sqrt[3]{\frac{\bar{\rho}_g}{\bar{\rho}_c}}\right).
\end{align}%
From the other hand, \eqref{ElastTeq} has exact solution in this spherical case when $R\gg b$:
\begin{align}
\delta r=&\alpha\frac{1+\nu}{1-\nu}T_1\frac{b^2}{a^2}\left(\frac{\sqrt\pi}{4}b\erf{\frac{a}{b}}-\frac{a}{2}e^{-\frac{a^2}{b^2}}\right),\\
\label{deltarandR}
\frac{\delta r}{\delta R}=&
\frac{\sqrt\pi b}
     {\sqrt\pi b\erf{\frac{a}{b}}-2ae^{-\frac{a^2}{b^2}}}
		\frac{R^2}{a^2},
\end{align}
letting us to estimate $T_1$ and $b$. We then search for a solution of \eqref{ElastTeq} in cylindrical coordinates for half-space in the form $\vec u=\vec u_1+{\rm grad}\phi$ where $\phi$ takes the right part of \eqref{ElastTeq}
\begin{align}
\Delta\phi&=\frac{1+\nu}{1-\nu}\alpha T
\end{align}
and gives the boundary for $\vec u_1$ problem. So the $\phi$ is a simple driven Poissonian solution of well known form and $\vec u_1$ problem can be treated as a boundary-driven halfspace problem solved in \cite{ll7en}. The result for the displacement field $\vec u$ should be taken on boundary and averaged over the profile of a Gaussian beam with radius $w$ to catch the measured displacement of the mirror.
\begin{align}
\label{ElemResponse}
\delta z_j(x_j,y_j,z_j)=\int u_z(x,y,0,x_j,y_j,z_j)\frac{2}{\pi w^2}e^{-2\frac{x^2+y^2}{w^2}}dS
\end{align}
After some calculations similar to \cite{BGVteC} we obtain the averaged surface response on crystallization of a bubble occuring at coordinates $x_j$, $y_j$, $z_j$
\begin{align}
\label{ThermoMethod}
\delta z_j(x_j,y_j,z_j)& =2\alpha T_1(1+\nu)\pi^{3/2}b^3\times\\
  &\quad \times \int e^{-k_\bot^2w^2/4}e^{-k^2b^2/4}e^{-i\vec k\vec r_j}\frac{k_\bot}{k^2}\frac{d^3k}{(2\pi)^3}\,. \nonumber
\end{align}
Note that $2\pi^{3/2}\alpha T_1 b^3(1+\nu)\approx \frac{6\pi(1-\nu)^2}{1+\nu}\left(1-\sqrt[3]{\frac{\bar\rho_g}{\bar\rho_c}}\right) a^3=\xi V_a$ due to \eqref{deltar}-\eqref{deltarandR}, eliminating thermodynamical parameters. Here we changed $4/3\pi a^3$ to $V_a$ -- the volume of a collapsing region, in attempt to generalize formulas to an arbitrary bubble geometry.

\begin{figure}[ht]
\center
\includegraphics[width=0.48\textwidth]{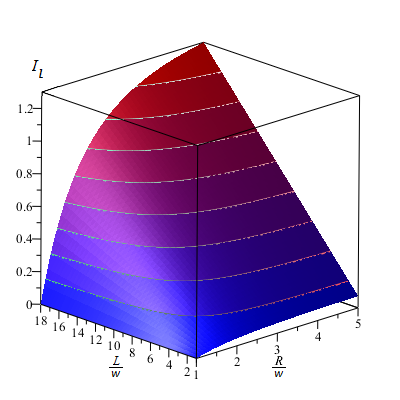}
\includegraphics[width=0.48\textwidth]{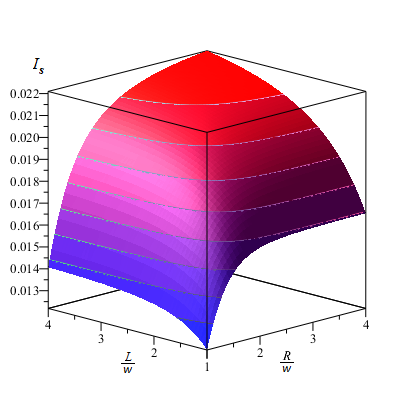}
\caption{Dimensionless integrals $I_l$ for the case of long cylinder (analytics that was used in \eqref{contrspeed}) and numerically calculated $I_S$ for a mirror (based on \eqref{ThermoMethod}).}
\label{Ifigs}
\end{figure}

The process of noise consists of discreet collapses
\begin{align}
\delta z(t)=\sum_j^{N(t)}H(t-\tau_j)\delta z_j(x_j,y_j,z_j)
\end{align}
where $x_j,y_j,z_j,\tau_j$ position and time of a collapse and $H(t)$ is a Heaviside step function.
For the stationary Gaussian process we obtain
\begin{align}
\langle\delta z(t)\rangle&=\lambda\int_V\delta z_j(x,y,z)dV\int_0^tH(t-\tau)d\tau= \nonumber\\
\label{AVlength}
  &\qquad =\lambda\xi V_a wI_l(R/w,L/w)t\\
  \label{S}
S_{\delta z}&=\lambda \int_V\delta z_j(x,y,z)^2 dV\tilde H(-\omega)\tilde H(\omega)=\nonumber\\
      &\qquad = \frac{\lambda \xi^2 V_a^2 I_S(R/w,L/w)}{w\omega^2},
\end{align}
where $\lambda$ is the process rate parameter -- number of events per second in unit volume, $I_l$ and $I_S$ are numerical values of underlying integrals, $R$ and $L$ -- radius and thickness of the mirror. The numerically calculated dimensionless integrals $I_l$ and $I_S$ are represented on Fig. \ref{Ifigs}.

In \cite{StabilitySilica} a contraction of a silica Fabri-Perot etalon was measured. Two mirrors with a diameter of about $w_e=0.66$ cm were connected with a $L_e=10$ cm long tube with outer diameter of $R_e=2$ cm. For this geometry \eqref{ElemResponse} should be changed as the measurement is equivalent to averaging over a ring and not a Gaussian spot. Furthermore the \cite{BGVteC} approach is not precise as it uses an assumption of infinite half-space, while here we need a long thin cylinder. To overcome this issue we use FEM modeling.
\begin{figure}
\includegraphics[width=0.48\textwidth]{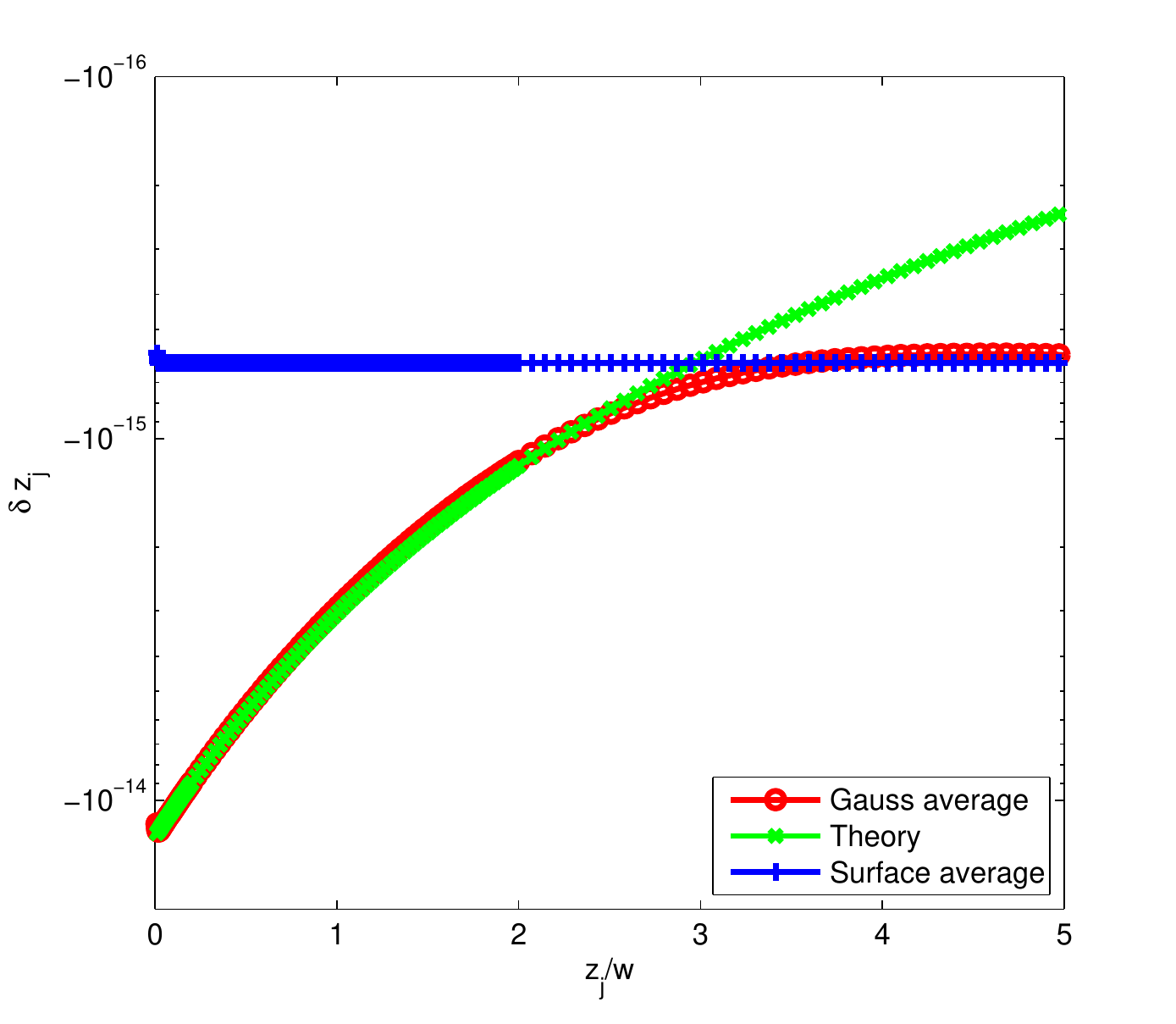}
\includegraphics[width=0.48\textwidth]{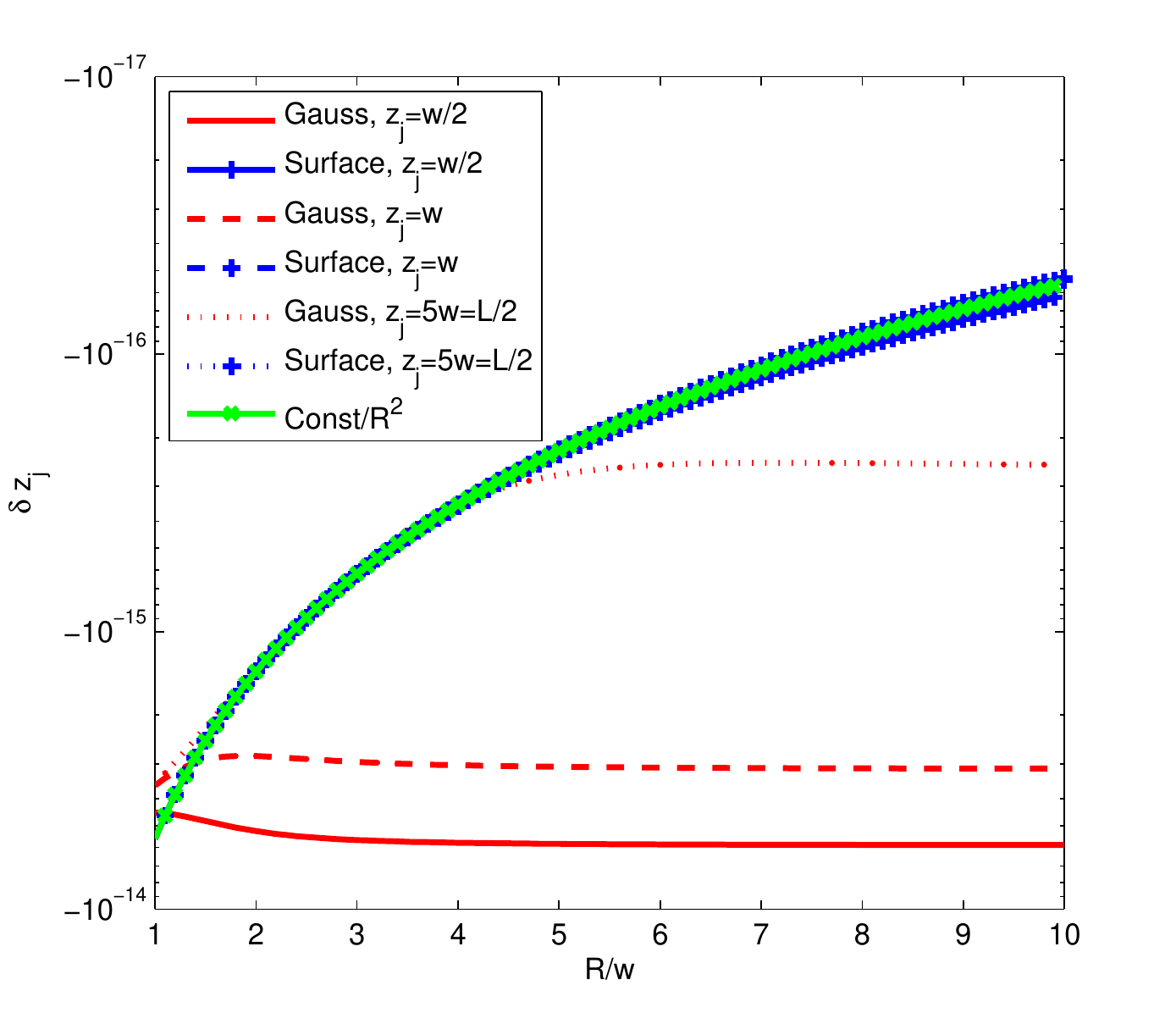}
\caption{Averaged $z$-displacement done by a bubble on the cylinder axis as a function of $z_j$ (top) and cylinder radius (bottom). The averaging is made with and without Gaussian function (red and blue-plus lines). The theory (green-crosses) line in the top figure is \eqref{ThermoMethod}. The $R$-dependence is taken for different bubble depth $z_j$ and is very close to ${\rm const}/R^2$ dependence (green-cross line) in all cases of surface-averaging.}
\label{avu(hc,R)}
\end{figure}
A cylinder with the above parameters was modeled using Comsol Multiphysics. Structural Mechanics module was used with two different problem formulations: a direct boundary load problem and a prescribed temperature problem (thermal expansion node under Linear Elastic Material node). The two solutions were found identical with respect to force normalization. The main results of the modeling are shown on Fig. \ref{avu(hc,R)}--\ref{avu(hc,rc)}. 
\begin{figure}
\includegraphics[width=0.48\textwidth]{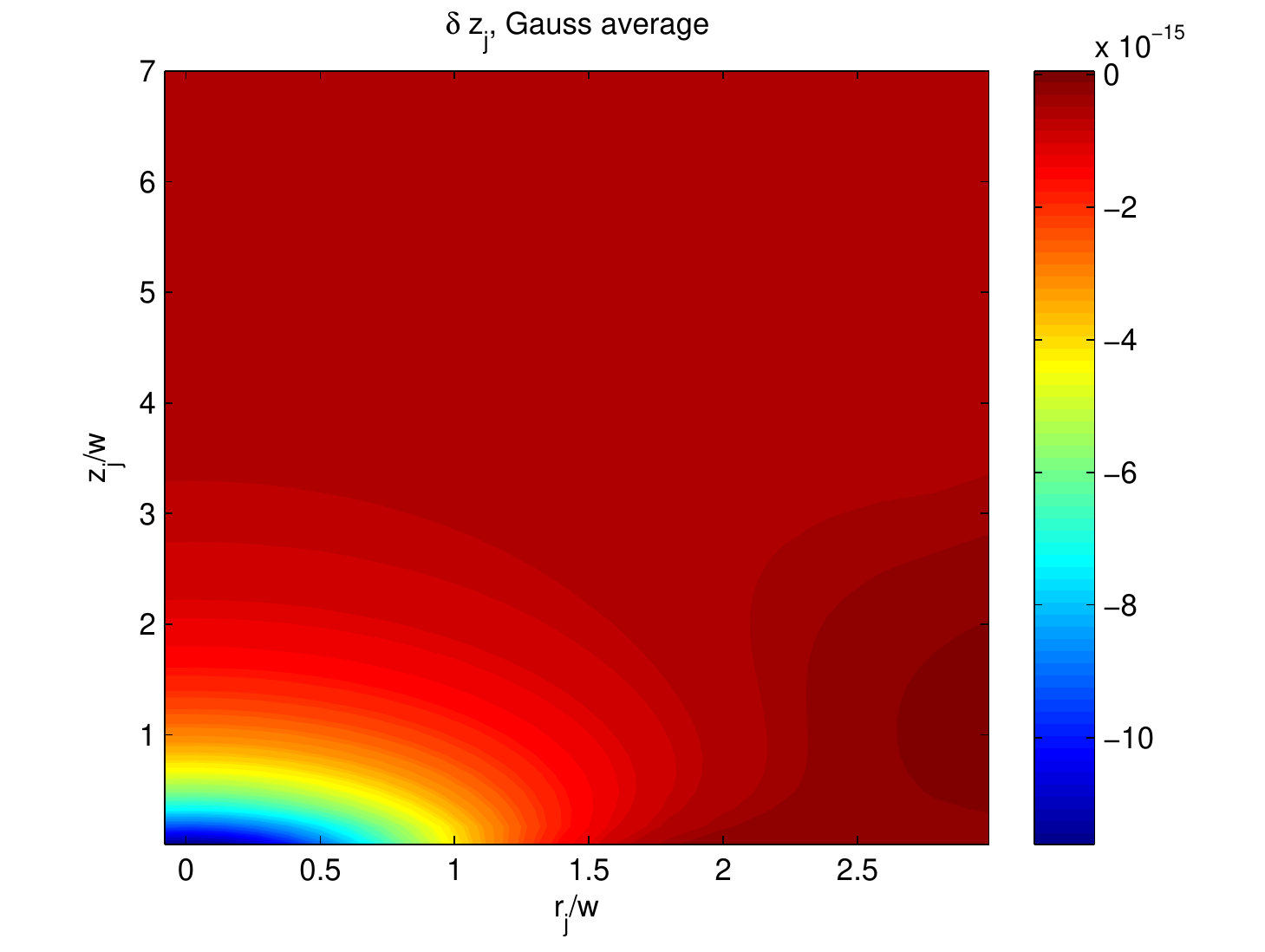}
\includegraphics[width=0.48\textwidth]{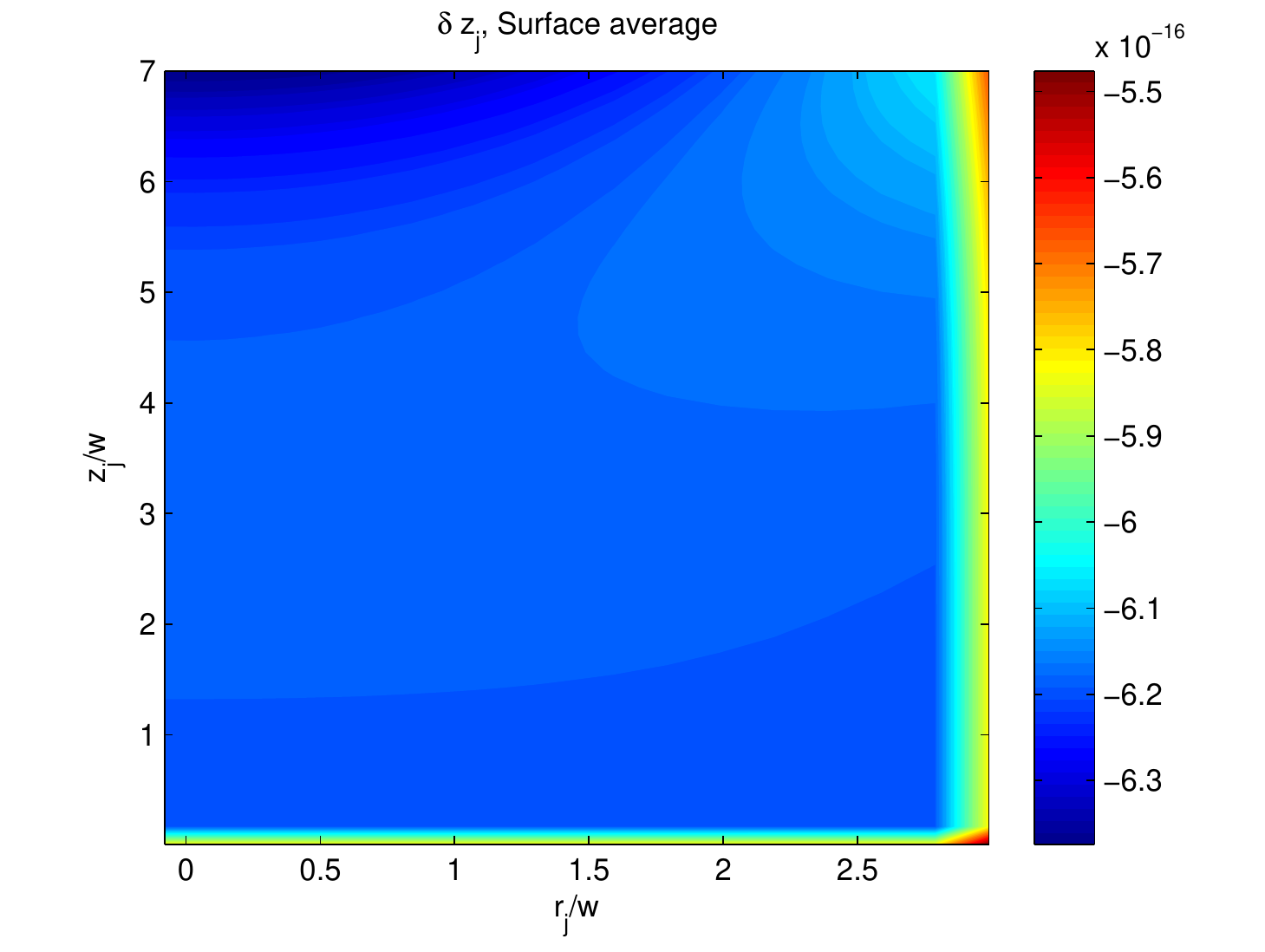}
\caption{Averaged $z$-displacement done by a bubble on the cylinder axis as a function of $r_j$ and $z_j$ for Gaussian averaging (top) and surface averaging (bottom).}
\label{avu(hc,rc)}
\end{figure}

From the simulations it follows that the whole region of a cylinder can be subdivided in two parts: 1) a sphere with the center in the center of a collapsing bubble, touching the closest surface of the cylinder and 2) the rest of the cylinder. The solution inside the sphere is close to the spherically symmetric solution \eqref{sphereForm}. Outside the sphere the solution is close to constant spherical field.
The averaged elementary response from one bubble in case of surface averaging is found to be practically constant with depth and varies less than by $0.8\%$ with offset from the axis. 

The averaged elementary response from one bubble in case of Gaussian averaging is in good agreement with the theory \eqref{ThermoMethod} till the depth of $2.5w$, and approaches the first modeling case after $3.6w$ (see fig. \ref{avu(hc,R)} top). That can be explained easily, having in mind that the cylinder radius here was $R=3w$. The idea is that the bubble ``feels'' only the closest boundary (the one that is touched by the earlier  mentioned sphere). For the depths less than $R$ (and close to the cylinder axis) the governing boundary is the front surface, making the problem similar to half-space and demonstrating appropriate transversal variations for Gaussian averaging and constant for surface averaging (see fig. \ref{avu(hc,rc)}). For greater depths the governing surface is the side surface, and the distance from it to the collapses near cylinder axis is constant, providong a constant response from depth.

In this way, for noise calculation we assume that \eqref{ThermoMethod} is valid until the depth of the bubble is smaller than $R$ and stays constant for larger depths with an error no larger than 13\%. However, as LIGO mirrors have $L\approx R$ we thus stay in half-space approximation for spectral density and do not need the $z>R$ extension. For the time constant determination we assume the elementary response from one bubble to be constant and equal to $\delta z_j(0,0,R)$. With these arguments the contraction and noise \eqref{ThermoMethod} can be found analytically. Then the limit of small $w$ should be taken to remove the Gaussian beam radius from expressions. It ensures that the response value at the depth $R$ is equal to the one of surface averaging case. Thus we get $\delta z_j(0,0,R)\propto R^{-2}$ which is shown as green line on Fig. \ref{avu(hc,R)} (right).
For the process parameter we obtain 
\begin{align}
\label{contrspeed}
\lambda=\frac{\dot{\langle\delta z\rangle}}{L_e}\frac{2}{V_a\xi}\frac{2R_e^2}{R_e^2-w_e^2}
\end{align}

We assume that the size of collapse is of the order of silica molecule $V_a\approx\frac{M_{\rm SiO_2}}{\bar\rho_{c}N_A}=45\times10^{-30}$ m$^3$ ($a\approx0.22$ nm), where $M_{\rm SiO_2}=60$ g/mol -- molar mass, $N_A$ -- Avogadro number.
From \cite{StabilitySilica} we get $\dot\epsilon_{zz}=-5.8\times10^{-15}$ per second and calculate $\lambda\approx1.54\times10^{16}$ events per second per m$^3$.
The resulting spectral density of devitrification noise at 100 Hz for $R=L=20$ cm and $w=6$ cm is
\begin{align}\label{sqrtS}
\sqrt{S_{\delta z}}&=\sqrt{\frac{\dot\epsilon_{zz}\xi V_a}{w\omega^2}\frac{L_e}{w_e}\frac{I_S(R/w,L/w)}{I_l(R_e/w_e,L_e/w_e)}}\nonumber\\
									&=6.31\times10^{-25}\, \text{ m/Hz}^{1/2},
\end{align}
which is $8000$ times smaller than the Brownian noise in substrates and coatings for LIGO mirrors \cite{MyThermalNoise,Harry}.

\section{Devitrification noise in string suspensions}

\begin{figure}[ht]
\includegraphics[width=0.48\textwidth]{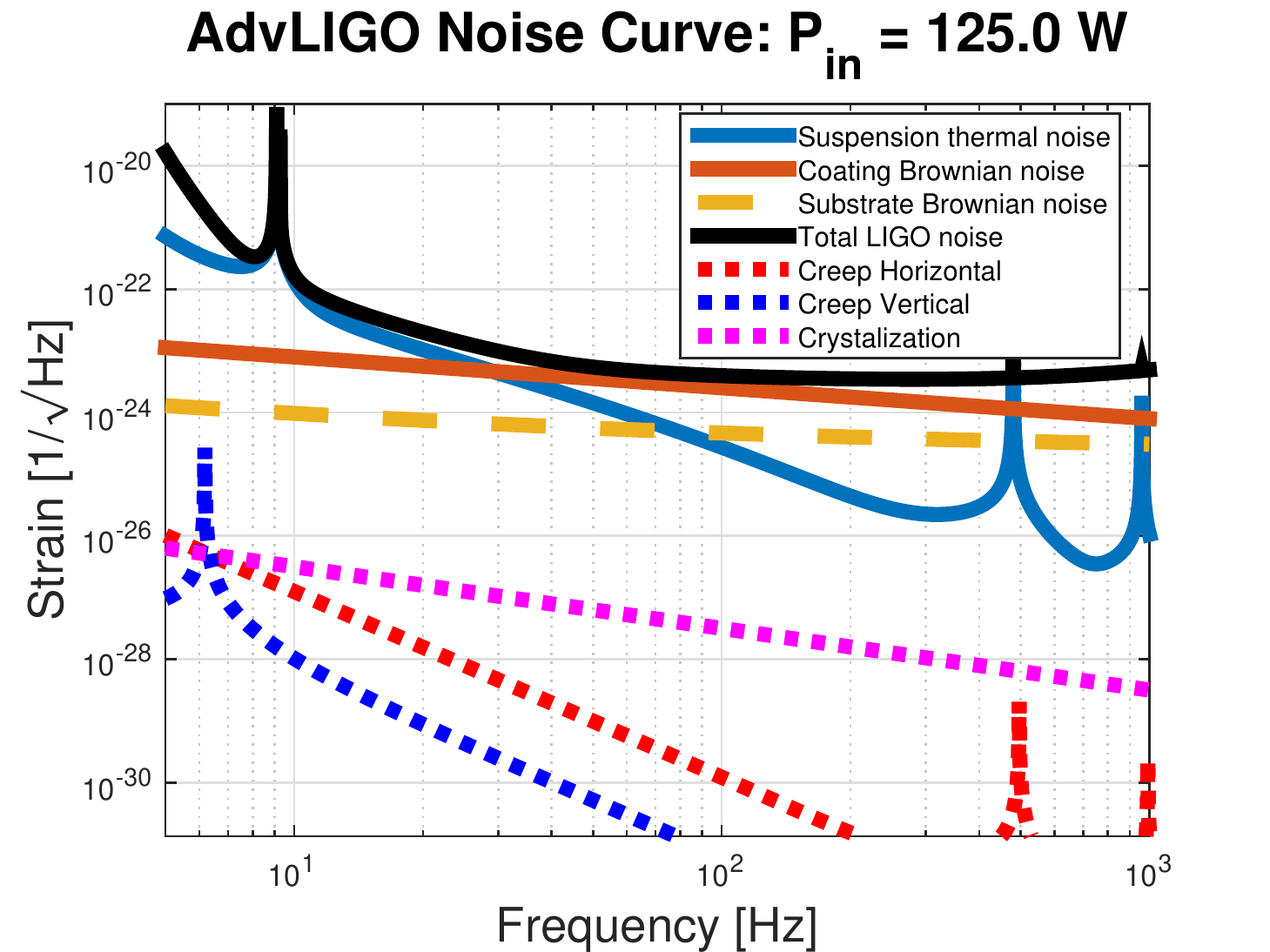}
\caption{Creep suspension noises from \cite{CreepNoise} using process parameter of devitrification \eqref{contrspeed} and devitrification noise \eqref{sqrtS} in mirrors together with other Advanced LIGO noises.}
\label{Sugar}
\end{figure}

Another estimate of event-based noise was made recently for suspension fibers by Yu. Levin \cite{CreepNoise}. He considered spontaneous discrete stress relaxation events (creep events) in suspension strings. However he also suffered from the lack of the process rate parameter and event volume values and thus could not obtain absolute noise values. One can speculate about the origin and direction of the creep events, but local reordering may be one of the sources. So we can use formulas (49) and (52) from \cite{CreepNoise} to estimate devitrification noise in suspensions, changing $R\langle V^2\rangle$ (Levin's notation) to $NV_s\lambda V_a^2$ (our notation), where $V_s=\pi r_sl_s$ -- the volume of the string and $N$ is the number of strings. Three devitrification noises together with existing LIGO noises are shown on Fig. \ref{Sugar}.

This naive estimate of suspension noise due to devitrification does not take into account complex multistage suspension system of LIGO. Furthermore the time parameter for the loaded case (mirror mass is about 40 kg) is probably smaller because the extension caused by massive mirrors opposes crystallization with contraction. Nevertheless we present at least the upper bound, which is already $2\times10^5$ times smaller then the Brownian suspension noise.

\section{Conclusion}

The main uncertainty of our estimates is the average radius of collapsing bubbles. Our initial idea to estimate it from the crystal-glass internal energy difference encountered  a serious problem as this energy is also not known exactly. Different estimates from literature give values varying by two-three times.

Note that spectral density \eqref{S} strongly depends on average radius of collapsing bubbles for constant $\lambda$: $S_{\delta z}\sim \lambda a^6$. However, taking in account that the value of $\lambda a^3$ is taken from contraction rate \eqref{contrspeed}, we get that effectively $\sqrt{S_{\delta z}} \sim  a^{3/2}$, i.e.  increase of the radius $a$ 10 times will increase the estimate {\eqref{sqrtS}} $30$ times.
It means that reliable knowledge of collapsing radius $a$ (as well as $\lambda$) is very important.

\acknowledgments
SPV and MLG acknowledge support from the Russian
Foundation for Basic Research (Grants No. 14-02-00399A
and No. 13-02-92441 in the frame of program ASPERA) and
National Science Foundation (Grant No. PHY-130586).

\bibliography{XNoise}

\end{document}